\documentclass[a4paper, 12pt, oneside]{article}
\usepackage[cp1251]{inputenc} 
\usepackage[english, russian]{babel} 
\usepackage{ graphics,amsfonts, amsmath,bm,amssymb, array, eucal}
\usepackage[dvips]{graphicx}
\usepackage[flushleft,small]{caption2} 

\makeatletter
\renewcommand{\@biblabel}[1]{#1.\hfill}

\newcommand{\const}{\mathop{\rm const\, }}

\makeatother {\renewcommand{\baselinestretch}{1.2}
\usepackage[usenames]{color}

\begin{document}
\newcommand{\mc}[1]{\mathcal{#1}}
\newcommand{\E}{\mc{E}}
\thispagestyle{empty}
\large

\renewcommand{\abstractname}{\,}
\renewcommand{\refname}{\begin{center} REFERENCES\end{center}}

 \begin{center}
\bf Longitudinal electric conductivity and dielectric permeability
in quantum plasma with variable frequency of collisions in Mermin' approach
\end{center}\medskip
\begin{center}
  \bf A. V. Latyshev\footnote{$avlatyshev@mail.ru$} and
  A. A. Yushkanov\footnote{$yushkanov@inbox.ru$}
\end{center}\medskip

\begin{center}
{\it Faculty of Physics and Mathematics,\\ Moscow State Regional
University, 105005,\\ Moscow, Radio str., 10--A}
\end{center}\medskip

\begin{abstract}
Formulas for longitudinal electric conductivity and dielectric permeability
in the quantum non-degenerate collisional plasma with the frequency of
collisions depending on momen\-tum in Mermin' approach are received.
The kinetic equation in momen\-tum space in relaxation approximation is
used.
It is shown that when Planck's constant tends to zero, the deduced
formula passes to the corresponding formula for classical plasma.
It is shown also that when frequency of collisions of particles of plasma
tends  to zero (plasma passes to collisionless one), the deduced formula
passes to the known Lindhard' formula received for collisionless
plasmas.
It is shown, that when frequency of collisions is a constant,
the deduced formula for dielectric permeability
passes in known Mermin' formula.


\end{abstract}

\begin{center}
{\bf Introduction}
\end{center}

In the known work of Mermin \cite{Mermin} on the basis of the
analysis of nonequilib\-rium density matrix in $\tau $--approximation
has been obtained expression for longitu\-di\-nal dielectric permeability of
quantum collisional plasmas for case of constant collision frequency of plasmas
particles.
Earlier in the work of Klimontovich and Silin \cite{Klim} and after that in
the work of Lindhard \cite{Lin} has been obtained
expression for longitudinal and transverse dielectric permeability of quantum
collisionless plasmas. By Kliewer and Fuchs \cite{Kliewer} it has been shown, that
direct generalisation of formulas of Lindhard on the case of collisional plasmas
is incorrectly. This lack for the longitudinal dielectric
permeability has been eliminated in the work of Mermin \cite{Mermin}.
Next in the work \cite{Das} has been given attempt to deduce Mermin's
formula.

For collisional plasmas correct formulas longitudinal and transverse
electric conductivity and dielectric permeability are received
accordingly in works \cite{Long} and \cite{Trans}. In these works
Wigner---Vlasov---Boltzmann kinetic equation
in relaxation approximation in coordinate space is used. In work
\cite{Trans2} the formula for transverse electric conductivity has been deduced
for quantum collisional plasmas with use of the kinetic equation
by Mermin' approach (in momentum space).

In the present article formulas for longitudinal electric conductivity
and dielectric permeability in the quantum non-degenerate collisional
plasma with the frequency of
collisions depending on an momen\-tum by Mermin' approach are received.
The kinetic equation in momen\-tum space in relaxation approximation is
used.
It is shown, that when Planck's constant tends to zero, the deduced
formula passes to the corresponding formula for classical plasma.
It is shown also, that when frequency of collisions of particles of plasma
tends  to zero (plasma passes to collisionless one), the deduced formula
passes to the known Lindhard' formula received for collisionless
plasmas. It is shown, that when frequency of collisions is a constant,
the deduced formula for dielectric permeability
passes in known Mermin' formula.

Now considerable interest to research of properties of quantum plasma proceeds
\cite{Manf}--\cite{Ropke}.

\begin{center}
  \bf 1. Kinetic Schr\"{o}dinger---Boltzmann equation for density matrix
\end{center}

Let the vector potential of an electromagnetic field is
harmonious, i.e. changes as
$
{\varphi}({\bf r},t)={\varphi}({\bf r})\exp(-i \omega t).
$
Relation between scalar potential and intensity of the electric
field it is given by expression
$
{\bf E}({\bf q})=-\nabla{\varphi}({\bf q}).
$
The equilibrium matrix of density looks like
$$
{\tilde \rho}=\dfrac{1}{\exp\dfrac{H-\mu}{k_BT}+1},\qquad \mu=\mu_0+\delta\mu.
$$

Here $T $ is the temperature, $k_B$ is the Boltzmann constant,
$\mu_0$ is the chemical potential of plasma in an equilibrium condition,
$\delta\mu $ is the correction to the chemical potential, caused
presence of variable electric field, $H $ is the Hamiltonian.

Hamiltonian looks like here $H=H_0+H_1$, where
$H_0={\bf p}^2/2m,\; H_1=e\varphi$.
Here $m, e$ are mass and charge of electron, ${\bf p}=\hbar{\bf k}$
is the electron momentum.

Let's designate an equilibrium matrix of density in absence of an external field
through ${\tilde \rho}_0$:
$$
{\tilde \rho}_0=\dfrac{1}{\exp \dfrac{H_0-\mu_0}{k_BT}+1}.
$$

Density matrix it is possible to present an equilibrium matrix of density
in the form
$$
{\tilde \rho}={\tilde \rho}_0+{\tilde \rho}_1.
$$

Here ${\tilde \rho}_1$ is the correction to the equilibrium matrix of density,
caused by presence of an electromagnetic field.

In linear approximation we receive
$$
[H, {\tilde \rho}\,]=[H_0, {\tilde \rho}_1]+[H_1,{\tilde\rho}_0],
$$
and
$$
[H, {\tilde \rho}\,]=0.
$$

Here $[H, {\tilde \rho}\,]=H {\tilde \rho}-{\tilde \rho}H$ is
the commutator.

Let's notice, that the vector $|{\bf k}\rangle $ is the eigen vector
of operators $H $ and $ \bf p $. Thus
$$
H|{\bf k}\rangle=\E_{\bf k}|{\bf k}\rangle,\quad
\langle{\bf k}|H=\E_{\bf k}\langle{\bf k}|, \quad
{\bf p}|{\bf k}\rangle=\hbar{\bf k}|{\bf k}\rangle, \quad
\langle{\bf k}|{\bf p}=\hbar{\bf k}\langle{\bf k}|.
$$

Let's notice, that for the operator $L $ the relationship is carried out
$$
\langle{\bf k}_1|L|{\bf k}_2\rangle=
\dfrac{1}{(2\pi)^3}\int\exp(-i{\bf k}_1{\bf r})L\exp(i{\bf k}_2{\bf
r})d{\bf r}.
$$

By means of this relation it is received
$$
\left\langle{\bf k}_1\left|[H_0, {\tilde \rho}_{1}]\right|{\bf k}_2
\right\rangle=-\left\langle{\bf k}_1\left|[H_1, {\tilde \rho}_{0}]\right|
{\bf k}_2\right\rangle.
$$

Let's write down this equality in the explicit form
$$
\left\langle{\bf k}_1\left|H_0{\tilde \rho}_1\right|{\bf k}_2
\right\rangle - \left\langle{\bf k}_1\left|{\tilde \rho}_1 H_0\right|{\bf k}_2
\right\rangle=-\left\langle{\bf k}_1\left|H_1 {\tilde \rho}_0\right|
{\bf k}_2\right\rangle+\left\langle{\bf k}_1\left|{\tilde \rho}_0 H_1 \right|
{\bf k}_2\right\rangle.
$$

From here we receive, that
$$
(\E_{{\bf k}_1}-\E_{{\bf k}_2})\langle {\bf k}_1|{\tilde \rho}_1|{\bf k}_2)
\rangle=
(f_{{\bf k}_1}-f_{{\bf k}_2})\langle{\bf k}_1|H_1|{\bf k}_2\rangle=
$$
$$
=e(f_{{\bf k}_1}-f_{{\bf k}_2})\langle{\bf k}_1|\varphi|{\bf k}_2\rangle.
$$
Here
$$
\E_{{\bf k}}=\dfrac{\hbar^2 {\bf k}^2}{2m},\qquad
f_{{\bf k}}=\dfrac{1}{\exp\dfrac{\E_{{\bf k}}-\mu_0}{k_BT}+1}.
$$

Considering, that
$$
\langle{\bf k}_1|\varphi|{\bf k}_2\rangle=\dfrac{1}{(2\pi)^3}
\int\exp(-i({\bf k}_1-{\bf k}_2){\bf r})
\varphi({\bf r})d{\bf r}=\varphi({\bf k}_1-{\bf k}_2),
$$
we receive
$$
(\E_{{\bf k}_1}-\E_{{\bf k}_2}){\tilde \rho}_{1}({\bf k}_1,{\bf k}_2)=
e(f_{{\bf k}_1}-f_{{\bf k}_2})\varphi({\bf k}_1-{\bf k}_2).
$$

The kinetic equation for the density matrix in $\tau$--approximation with
constant frequency of collisions looks like
$$
i\hbar \dfrac{\partial \rho}{\partial t}=[H,\rho]+\dfrac{i\hbar}{\tau}({\tilde
\rho}-\rho),
$$
or
$$
\dfrac{\partial \rho}{\partial t}=-\dfrac{i}{\hbar}[H,\rho]+\nu({\tilde
\rho}-\rho).
\eqno{(1.1)}
$$

Here $\tau$ is the average time of free electrons path,
$ \nu=1/\tau $ is the frequency of collisions.

Generally frequency of collisions $ \nu $ should depend from
electron mo\-men\-tum $ {\bf p} $ (or a wave vector {\bf k}): $ \nu =\nu ({\bf
k}) $.

Considering the requirement Hermitian character the equation (1.1) on
the density matrix it is necessary to rewrite in the form
$$
\dfrac{\partial \rho}{\partial t}=
-\dfrac{i}{\hbar}[H,\rho]+\dfrac{\nu({\bf k})}{2}
 ({\tilde \rho}-\rho)+({\tilde \rho}-\rho)\dfrac{\nu({\bf k})}{2}.
 \eqno{(1.2)}
$$

In linear approximation  the density matrix we will search in the form
$$
\rho={\tilde \rho}_{0}+{\rho}_{1}.
$$

Then in linear approach the equation (1.2) looks like
$$
i\hbar\dfrac{\partial \rho_1}{\partial t}=H_0\rho_1-\rho_1H_0+H_1\rho_0-
\rho_0 H_1+$$$$+i\hbar\dfrac{\nu({\bf k})}{2}
 ({\tilde \rho_1}-\rho_1)+i \hbar({\tilde \rho_1}-\rho_1)\dfrac{\nu({\bf k})}{2}.
\eqno{(1.3)}
$$

Let's consider, that $ \rho_{1}\sim \exp(-i\omega t) $.
From here for $ {\rho}_{1} $ we receive the relation
$$
\hbar \omega\langle{\bf k}_1|\rho_{1}|{\bf k}_2\rangle=
(\E_{{\bf k}_1}-\E_{{\bf k}_2})
\langle{\bf k}_1|\rho_{1}|{\bf k}_2\rangle-
e(f_{{\bf k}_1}-f_{{\bf k}_2}){\varphi}({\bf k}_1-{\bf k}_2)+
$$
$$+\dfrac{i\hbar\nu({\bf k}_1)}{2}
\langle{\bf k}_1|{\tilde \rho_1}- \rho_1|{\bf k}_2\rangle+
\langle{\bf k}_1|{\tilde \rho_1}- \rho_1|{\bf k}_2
\rangle\dfrac{i\hbar\nu({\bf k}_2) }{2},
$$
or, having designated
$$
\bar \nu(\mathbf{k}_1,\mathbf{k}_2)=\dfrac{\nu({\bf k}_1)+\nu({\bf k}_2)}{2},
$$
Let's rewrite the previous equation in the form
$$
\hbar \omega\langle{\bf k}_1|\rho_{1}|{\bf k}_2\rangle=
(\E_{{\bf k}_1}-\E_{{\bf k}_2})
\langle{\bf k}_1|\rho_{1}|{\bf k}_2\rangle
-e(f_{{\bf k}_1}-f_{{\bf k}_2}){\varphi}({\bf k}_1-{\bf k}_2)+
$$
$$
+i\hbar\bar \nu(\mathbf{k}_1,\mathbf{k}_2)
\langle{\bf k}_1|{\tilde \rho}_{1}-\rho_{1}|{\bf k}_2\rangle.
\eqno{(1.4)}
$$

From equation (1.4) we receive
$$
\langle{\bf k}_1|\rho_{1}|{\bf k}_2\rangle=\rho_1(\mathbf{k}_1,{\bf k}_2)=
-e\dfrac{f_{{\bf k}_1}-f_{{\bf k}_2}}{\E_{{\bf k}_2}-\E_{{\bf k}_1}+
\hbar (\omega+i\bar \nu(\mathbf{k}_1,\mathbf{k}_2))}\varphi({\bf k}_1-{\bf k}_2)+
$$
$$
+\dfrac{i\hbar \bar \nu \langle{\bf k}_1|\tilde\rho_{1}|{\bf k}_2\rangle}
{\E_{{\bf k}_2}-\E_{{\bf k}_1}+\hbar (\omega+i\bar \nu(\mathbf{k}_1,\mathbf{k}_2))}.
\eqno{(1.5)}
$$

The relation (1.5) represents the solution of linear
Schr\"{o}din\-ger---Boltz\-mann equation, expressed through perturbation to
equilibrium mat\-rix of density
$\tilde{\rho}_1({\bf k}_1-{\bf k}_2)=
\langle{\bf k}_1|\tilde\rho_{1}|{\bf k}_2\rangle$.
Let's find this perturbation.

Let's take advantage of an obvious relation
$$
[H-\mu,{\tilde \rho}]=0.
$$
In linear approximation from here it is received
$$
[H_0-\mu_0,{\tilde \rho}_{1}]+[H_1-\delta\mu,{\tilde \rho}_{0}]=0.
$$

Transforming the first commutator, from here we receive:
$$
[H_0,{\tilde \rho}_{1}]=[\delta\mu-H_1,{\tilde \rho}_{0}],
$$
or
$$
[H_0,{\tilde \rho}_{1}]=[\delta\mu-e\varphi,{\tilde \rho}_{0}].
$$

Let's designate now
$$
\delta\mu_*=\delta\mu-e\varphi.
$$

Then the previous equality write down in the form
$$
[H_0,{\tilde \rho}_{1}]=[\delta\mu_*,{\tilde \rho}_{0}].
$$

From here we receive that
$$
(\E_{{\bf k}_1}-\E_{{\bf k}_2})
\langle{\bf k}_1|{\tilde\rho_{1}}|{\bf k}_2\rangle
=-(f_{{\bf k}_1}-f_{{\bf k}_2})\delta\mu_*({\bf k}_1-{\bf k}_2),
$$
from which
$$
\langle{\bf k}_1|{\tilde\rho_{1}}|{\bf k}_2\rangle=-
\dfrac{f_{{\bf k}_1}-f_{{\bf k}_2}}
{\E_{{\bf k}_1}-\E_{{\bf k}_2}}
\langle{\bf k}_1|\delta \mu_*|{\bf k}_2\rangle,
\eqno{(1.6)}
$$
or, that all the same,
$$
\tilde{\rho}_1({\bf k}_1,{\bf k}_2)=-\dfrac{f_{{\bf k}_1}-f_{{\bf k}_2}}
{\E_{{\bf k}_1}-\E_{{\bf k}_2}}\delta\mu_*({\bf k}_1-{\bf k}_2).
$$

We have received perturbation to the equilibrium matrix of the density, expressed
through perturbation of chemical potential. The last we will find from the
preservation law of numerical density.

We put next
$${\bf k}_1={\bf k}+\dfrac{\mathbf{k}}{2},\qquad
{\bf q}_2={\bf k}-\dfrac{\mathbf{q}}{2}
$$
and rewrite in this terms equations
(1.4), (1.5) и (1.6). We receive following equalities
$$
\hbar \omega\Big \langle {\bf k}+\dfrac{{\bf q}}{2} \Big|\rho_1\Big|{\bf k}-
\dfrac{{\bf q}}{2}\Big\rangle=
\Big(\E_{{\bf k}+{\bf q}/2}-\E_{{\bf k}-{\bf q}/2}\Big)
\Big \langle {\bf k}+\dfrac{{\bf q}}{2} \Big|\rho_1\Big|{\bf k}-
\dfrac{{\bf q}}{2}\Big\rangle-
$$\medskip
$$
-e\varphi({\bf q})(f_{{\bf k}+{\bf q}/2}-
f_{{\bf k}-{\bf q}/2})+i\bar\nu\hbar
\Big \langle {\bf k}+\dfrac{{\bf q}}{2} \Big|\tilde \rho_1-\rho_1\Big|{\bf k}-
\dfrac{{\bf q}}{2}\Big\rangle,
\eqno{(1.4')}
$$\\
$$
 \Big\langle {\bf k}+\dfrac{{\bf q}}{2} \Big|\rho_1\Big|{\bf k}-
\dfrac{{\bf q}}{2}\Big\rangle \equiv \rho_1({\bf q})=-
\dfrac{e\varphi({\bf q})(f_{{\bf k}+{\bf q}/2}-
f_{{\bf k}-{\bf q}/2})}{\E_{{\bf k}-{\bf q}/2}-\E_{{\bf k}+{\bf q}/2}+\hbar \omega+
i\bar\nu \hbar}+
$$ \medskip
$$
+\dfrac{i \bar\nu \hbar  \Big\langle {\bf k}+\dfrac{{\bf q}}{2} \Big|\tilde\rho_1
\Big|{\bf k}- \dfrac{{\bf q}}{2}\Big\rangle}
{\E_{{\bf k}-{\bf q}/2}-\E_{{\bf k}+{\bf q}/2}+\hbar \omega+i\bar\nu \hbar},
\eqno{(1.5')}
$$\medskip
and
$$
 \Big\langle {\bf k}+\dfrac{{\bf q}}{2} \Big|\tilde\rho_1\Big|{\bf k}-
\dfrac{{\bf q}}{2}\Big\rangle=- \dfrac{f_{{\bf k}+{\bf q}/2}-
f_{{\bf k}-{\bf q}/2})}{\E_{{\bf k}+{\bf q}/2}-\E_{{\bf k}-{\bf q}/2}}
 \Big\langle {\bf k}+\dfrac{{\bf q}}{2} \Big|\delta \mu_*\Big|{\bf k}-
\dfrac{{\bf q}}{2}\Big\rangle.
\eqno{(1.6')}
$$\medskip

In this equalities (1.4')--(1.6') the designation
$$
\bar \nu=\bar\nu({\bf k,q})=
\dfrac{\nu\Big(\mathbf{k}+\dfrac{\mathbf{q}}{2}\Big)+
\nu\Big(\mathbf{k}-\dfrac{\mathbf{q}}{2}\Big)}{2}
$$
is used.

\begin{center}
\bf 2. Perturbation of chemical potential
\end{center}

The quantity $\delta\mu$ (or $\delta\mu_*$) is responsible for the local
preservation of number of particles (electrons). This local law
preservation looks like \cite{Mermin}
$$
\omega\delta n({\bf q}, \omega,\bar \nu)=
{\bf q}\delta{\bf j}({\bf q}, \omega, \bar \nu).
\eqno{(2.1)}
$$

In equation (2.1) $\delta n({\bf q},\omega,\bar \nu)$,
$\delta{\bf j}({\bf q},\omega,\bar \nu)={\bf j}({\bf q},\omega,\bar \nu)$
are сhange of concentration and stream density of electrons under action
electric field, and
$$
\delta n({\bf q}, \omega,\bar \nu)=\int \dfrac{d{\bf k}}{4\pi^3}
\left\langle{\bf k}+\frac{\mathbf{q}}{2}\Big|\rho_1\Big|{\bf k}-
\frac{\mathbf{q}}{2}\right\rangle,
$$
$$
\delta {\bf j}({\bf q}, \omega,\bar \nu)=
\int \dfrac{\hbar {\bf k}d{\bf k}}{4\pi^3m}
\left\langle{\bf k}+\frac{\bf q}{2}\Big|\rho_1\Big|{\bf k}-\frac{\bf q}{2}
\right\rangle.
$$

From equation on the matrix of density follows that
$$
\int \dfrac{\hbar \omega d{\bf k}}{4\pi^3}
\left\langle{\bf k}+
\frac{\bf q}{2}\Big|\rho_1\Big|{\bf k}-\frac{\bf q}{2}\right\rangle
=\hspace{7cm}
$$
$$
=\int \dfrac{ d{\bf k}}{4\pi^3}
\big(\E_{{\bf k}+{\bf q}/{2}}-\E_{{\bf k}-{\bf q}/{2}}\big)
\left\langle{\bf k}+\frac{\bf q}{2}\Big|\rho_1\Big|{\bf k}-
\frac{\bf q}{2}\right\rangle+
$$
$$
-e{\varphi}({\bf q})\int \dfrac{ d{\bf k}}{4\pi^3}
\Big(f_{{\bf k}+{\bf q}/{2}}-f_{{\bf k}-{\bf q}/{2}}\Big)=
$$
$$\hspace{4cm}
=i\hbar\int \dfrac{ d{\bf k}}{4\pi^3}\bar\nu({\bf k,q})
\left\langle{\bf k}+\frac{\bf q}{2}\Big|{\tilde \rho}_1-
\rho_1\Big|{\bf k}-\frac{\bf q}{2}\right\rangle.
$$

In this equalities  the designation
$$
\bar \nu=\bar\nu({\bf k,q})=
\dfrac{\nu\Big(\mathbf{k}+\dfrac{\mathbf{q}}{2}\Big)+
\nu\Big(\mathbf{k}-\dfrac{\mathbf{q}}{2}\Big)}{2},
$$
is used and  conditions
$$
\E_{{\bf k}+{\bf q}/{2}}-\E_{{\bf k}-{\bf q}/{2}}=
\dfrac{\hbar^2 }{m}{\bf k}{\bf q},
\eqno{(2.2)}
$$
and
$$
\int \dfrac{ d{\bf k}}{4\pi^3}\Big(f_{{\bf k}-{\bf q}/{2}}-
f_{{\bf k}+{\bf q}/{2}}\Big)=0
$$
are used.

Therefore last equality can be rewritten in the form
$$
\hbar\int \dfrac{ d{\bf k}}{4\pi^3}\Big(\omega
\left\langle{\bf k}+\dfrac{\mathbf{q}}{2}\Big|\rho_1\Big|{\bf k}-
\dfrac{\mathbf{q}}{2}\right\rangle-\dfrac{\hbar {\bf k}{\bf q}}{m}
\left\langle{\bf k}+\dfrac{\mathbf{q}}{2}\Big|\rho_1\Big|{\bf k}-
\dfrac{\mathbf{q}}{2}\right\rangle\Big)=
$$
$$
=i\hbar\int \dfrac{ d{\bf k}}{4\pi^3}\bar\nu(\mathbf{k,q})
\left\langle{\bf k}+\dfrac{\mathbf{q}}{2}\Big|{\tilde \rho}_1-
\rho_1\Big|{\bf k}-\dfrac{\mathbf{q}}{2}\right\rangle.
$$

Expression according to previous is equal to zero in the left part of
this relation, i.e.
$$
\hbar\omega\delta n({\bf q},\omega,\bar \nu)-
\hbar{\bf q}\delta{\bf j}({\bf q},\omega,\bar \nu)=0.
$$

The last is true owing to the law of local preservation of number
particles. From here follows, that
$$
\int \dfrac{ d{\bf k}}{4\pi^3}\bar \nu(\mathbf{k,q})
\left\langle{\bf k}+\dfrac{\mathbf{q}}{2}\Big|{\tilde \rho}_1-
\rho_1\Big|{\bf k}-\dfrac{\mathbf{q}}{2}\right\rangle=0.
$$
This equality is equivalent to the following
$$
\int \dfrac{ d{\bf k}}{4\pi^3}\bar\nu(\mathbf{k,q})
\left\langle{\bf k}+\dfrac{\mathbf{q}}{2}\Big|{\tilde\rho}_1\Big|{\bf k}-
\dfrac{\mathbf{q}}{2}\right\rangle=
$$
$$
=\int \dfrac{ d{\bf k}}{4\pi^3}\bar \nu(\mathbf{k,q})
\left\langle{\bf k}+\dfrac{\mathbf{q}}{2}\Big|
\rho_1\Big|{\bf k}-\dfrac{\mathbf{q}}{2}\right\rangle.
$$

Considering earlier received expression (1.4) for
$\langle{\bf k}_1|\tilde\rho_1|{\bf k}_2\rangle$, we have
$$
\delta\mu_*({\bf q},\bar \nu)\int \dfrac{ d{\bf k}}{4\pi^3}
\bar \nu(\mathbf{k,q})\dfrac{f_{{\bf k}+{\mathbf{q}}/{2}}-
f_{{\bf k}-{\mathbf{q}}/{2}}}
{\E_{{\bf k}-{\bf q}/2}-\E_{{\bf k}+{\bf q}/2}}=
$$
$$
=\int \dfrac{ d{\bf k}}{4\pi^3}\bar \nu(\mathbf{k,q})
\left\langle{\bf k}+\dfrac{\mathbf{q}}{2}\Big|
\rho_1\Big|{\bf k}-\dfrac{\mathbf{q}}{2}\right\rangle.
$$

Thus for perturbation quantity $ \delta\mu_*({\bf q}) $ it is received
$$
\delta\mu_*({\bf q},\omega,\bar \nu)=\dfrac{1}{B_{\bar \nu}({\bf q},0)}\int
\dfrac{ d{\bf k}}{4\pi^3}\bar \nu(\mathbf{k,q})
\left\langle{\bf k}+\dfrac{\mathbf{q}}{2}\Big|
\rho_1\Big|{\bf k}-\dfrac{\mathbf{q}}{2}\right\rangle.
\eqno{(2.3)}
$$

Here the following designation is accepted
$$
B_{\bar \nu}({\bf q},0)=
\int \dfrac{ d{\bf k}}{4\pi^3}
\bar \nu(\mathbf{k,q})
\dfrac{f_{{\bf k}+{\mathbf{q}}/{2}}-f_{{\bf k}-{\mathbf{q}}/{2}}}
{\E_{{\bf k}-{\bf q}/2}-\E_{{\bf k}+{\bf q}/2}}.
$$

From equation  $(1.3)$ we receive
$$
\Big[\E_{{\bf k}_2}-\E_{{\bf k}_1}+
\hbar \big(\omega+i\bar\nu(\mathbf{k}_1,\mathbf{k}_2)\big)\Big]
\langle{\bf k}_1|\rho_1|{\bf k}_2\rangle=
$$
$$
=-e(f_{{\bf k}_1}-f_{{\bf k}_2}){\varphi}({\bf k}_1-{\bf k}_2)
+i\hbar\bar\nu({\bf k}_1,{\bf k}_2)\langle{\bf k}_1|{\tilde \rho}_1|{\bf
k}_2\rangle.
$$
Last component in this equality we will replace according to (1.4).
We receive, that
$$
\Big[\E_{{\bf k}_2}-\E_{{\bf k}_1}+
\hbar \big(\omega+i\bar\nu(\mathbf{k}_1,\mathbf{k}_2)\big)\Big]
\langle{\bf k}_1|\rho_1|{\bf k}_2\rangle=
$$
$$
=-e(f_{{\bf k}_1}-f_{{\bf k}_2}){\varphi}({\bf k}_1-{\bf k}_2)-
$$
$$
-i\hbar \bar\nu({\bf k}_1,{\bf k}_2)
\dfrac{f_{{\bf k}_1}-f_{{\bf k}_2}}{\E_{{\bf k}_1}-\E_{{\bf k}_2}}
\delta\mu_*({\bf k}_1-{\bf k}_2,\omega,\bar \nu).
$$\medskip

From this equation we obtain expression for
$\langle{\bf k}_1|\rho_1|{\bf k}_2\rangle$:
$$
\langle{\bf k}_1|\rho_1|{\bf k}_2\rangle=-
\dfrac{e(f_{{\bf k}_1}-f_{{\bf k}_2}){\varphi}({\bf k}_1-{\bf k}_2)}
{\E_{{\bf k}_2}-\E_{{\bf k}_1}+\hbar (\omega+i\bar\nu({\bf k}_1,{\bf k}_2))}+
$$
$$
+i\hbar\bar\nu({\bf k}_1,{\bf k}_2)
\dfrac{f_{{\bf k}_1}-f_{{\bf k}_2}}{\E_{{\bf k}_2}-\E_{{\bf k}_1}}
\dfrac{\delta\mu_*({\bf k}_1-{\bf k}_2,\omega,\bar \nu)}
{\E_{{\bf k}_2}-\E_{{\bf k}_1}+\hbar (\omega+i\bar\nu({\bf k}_1,{\bf k}_2))},
\eqno{(2.4)}
$$ \medskip
or, after decomposition on partial fractions,
$$
\langle{\bf k}_1|\rho_1|{\bf k}_2\rangle=-
\dfrac{e(f_{{\bf k}_1}-f_{{\bf k}_2}){\varphi}({\bf k}_1-{\bf k}_2)}
{\E_{{\bf k}_2}-\E_{{\bf k}_1}+\hbar(\omega+i\bar\nu({\bf k}_1,{\bf k}_2))}+
$$
$$
+\dfrac{i\bar\nu({\bf k}_1,{\bf k}_2)}{\omega+i\bar \nu({\bf k}_1,{\bf k}_2)}
\cdot
\dfrac{f_{{\bf k}_1}-f_{{\bf k}_2}}{\E_{{\bf k}_2}-\E_{{\bf k}_1}}
\delta\mu_*({\bf k}_1-{\bf k}_2,\omega,\bar \nu)-
$$
$$
-
\dfrac{i\bar\nu({\bf k}_1,{\bf k}_2)}{\omega+i\bar \nu({\bf k}_1,{\bf k}_2)}
\cdot\dfrac{(f_{{\bf k}_1}-f_{{\bf k}_2})
\delta\mu_*({\bf k}_1-{\bf k}_2,\omega,\bar \nu)}
{\E_{{\bf k}_2}-\E_{{\bf k}_1}+\hbar(\omega+i\bar\nu({\bf k}_1,{\bf k}_2))},
\eqno{(2.4')}
$$ \medskip

Passing to variables $ {\bf k} $ and $ {\bf q} $, from here we receive
$$
\Big\langle{\bf k}+\dfrac{{\bf q}}{2}\Big|\rho_1\Big|{\bf k}-
\dfrac{{\bf q}}{2}\Big\rangle=-
\dfrac{e(f_{{\bf k+q}/2}-f_{{\bf k-q}/2}){\varphi}({\bf q})}
{\E_{{\bf k-q}/2}-\E_{{\bf k+q}/2}+\hbar(\omega+i\bar\nu({\bf k,q}))}+
$$
$$
+\dfrac{i\bar\nu({\bf k,q})}{\omega+i\bar \nu({\bf k,q})}\cdot
\dfrac{f_{{\bf k+q}/2}-f_{{\bf k-q}/2}}{\E_{{\bf k-q}/2}-\E_{{\bf k+q}/2}}
\delta\mu_*({\bf q},\omega,\bar \nu)-
$$
$$-
\dfrac{i\bar\nu({\bf k,q})}{\omega+i\bar \nu({\bf k,q})}\cdot
\dfrac{(f_{{\bf k+q}/2}-f_{{\bf k-q}/2})\delta\mu_*({\bf q},\omega,\bar \nu)}
{\E_{{\bf k-q}/2}-\E_{{\bf k+q}/2}+\hbar(\omega+i\bar\nu({\bf k,q}))}.
\eqno{(2.4'')}
$$\medskip

Let's substitute expression (2.4") in the formula for perturbation of chemical
potential (2.3). On this way for perturbation it is received the
following expression
$$
\delta\mu_*({\bf q},\omega,\bar \nu)=-e\varphi({\bf q})
\alpha_{\omega,\bar\nu}(\mathbf{q}).
\eqno{(2.5)}
$$

Here
$$
\alpha_{\omega,\bar\nu}(\mathbf{q})=
\dfrac{B_{\bar\nu}({\bf q},\omega+i\bar \nu)}{B_{\bar\nu}({\bf q},0)-
iB_{\omega,\bar\nu}({\bf q},0)+iB_{\omega,\bar\nu}({\bf q},\omega+i\bar \nu)},
$$ \medskip

$$
B_{\bar\nu}({\bf q},\omega+i\bar \nu)=
\int \dfrac{ d{\bf k}}{4\pi^3}
\bar \nu(\mathbf{k,q})
\dfrac{f_{{\bf k}+{\mathbf{q}}/{2}}-f_{{\bf k}-{\mathbf{q}}/{2}}}
{\E_{{\bf k}-{\bf q}/2}-\E_{{\bf k}+{\bf q}/2}+
\hbar(\omega+i \bar\nu({\bf k,q}))},
$$\medskip
$$
B_{\bar\nu}({\bf q},0)=
\int \dfrac{ d{\bf k}}{4\pi^3}
\bar \nu(\mathbf{k,q})
\dfrac{f_{{\bf k}+{\mathbf{q}}/{2}}-f_{{\bf k}-{\mathbf{q}}/{2}}}
{\E_{{\bf k}-{\bf q}/2}-\E_{{\bf k}+{\bf q}/2}},
$$\medskip
$$
B_{\omega,\bar\nu}({\bf q},0)=
\int \dfrac{ d{\bf k}}{4\pi^3}
\dfrac{f_{{\bf k}+{\mathbf{q}}/{2}}-f_{{\bf k}-{\mathbf{q}}/{2}}}
{\E_{{\bf k}-{\bf q}/2}-\E_{{\bf k}+{\bf q}/2}}\cdot \dfrac{\bar \nu^2({\bf k,q})}
{\omega+i\bar \nu({\bf k,q})},
$$\medskip
$$
B_{\omega,\bar\nu}({\bf q},\omega+i\bar \nu)=
\int \dfrac{ d{\bf k}}{4\pi^3}
\dfrac{f_{{\bf k}+{\mathbf{q}}/{2}}-f_{{\bf k}-{\mathbf{q}}/{2}}}
{\E_{{\bf k}-{\bf q}/2}-\E_{{\bf k}+{\bf q}/2}+
\hbar(\omega+i \bar\nu({\bf k,q}))}\cdot \dfrac{\bar \nu^2({\bf k,q})}
{\omega+i\bar \nu({\bf k,q})}.
$$\medskip

Let's write out a special case of the formula (2.5), when frequency of
collisions of particles of plasma it is constant:
$ \bar \nu ({\bf k, q}) \equiv \nu\equiv\const $. In this case perturbation
of chemical potential in Mermin's designations it is equal
$$
\delta\mu_*({\bf q},\omega,\nu)=-e\varphi({\bf q})\alpha(\omega,\nu).
\eqno{(2.6)}
$$

Here
$$
\alpha(\omega,\nu)=\dfrac{(\omega+i \nu)B({\bf q},\omega+i\nu)}
{\omega B({\bf q},0)+i \nu B({\bf q},\omega+i\nu)}.
$$ \medskip
$$
B({\bf q},0)=
\int \dfrac{ d{\bf k}}{4\pi^3}
\dfrac{f_{{\bf k}+{\mathbf{q}}/{2}}-f_{{\bf k}-{\mathbf{q}}/{2}}}
{\E_{{\bf k}-{\bf q}/2}-\E_{{\bf k}+{\bf q}/2}},
$$\medskip
$$
B({\bf q},\omega+i\nu)=
\int \dfrac{ d{\bf k}}{4\pi^3}
\dfrac{f_{{\bf k}+{\mathbf{q}}/{2}}-f_{{\bf k}-{\mathbf{q}}/{2}}}
{\E_{{\bf k}-{\bf q}/2}-\E_{{\bf k}+{\bf q}/2}+\hbar(\omega+i\nu)}.
$$ \medskip

\begin{center}
\bf 3. Electric conductivity and dielectric permeability
\end{center}

Let's substitute (2.5) in (2.4") and in the received expression we will
result similar members.
It is as a result received the following expression
$$
\Big\langle {\bf k}+\dfrac{{\bf q}}{2}\Big| \rho_1\Big|{\bf k}-
\dfrac{{\bf q}}{2}\Big\rangle=
$$
$$
-e \varphi({\bf q})\Bigg[
\dfrac{f_{{\bf k}+{\mathbf{q}}/{2}}-f_{{\bf k}-{\mathbf{q}}/{2}}}
{\E_{{\bf k}-{\bf q}/2}-\E_{{\bf k}+{\bf q}/2}+
\hbar(\omega+i\bar \nu({\bf k,q}))}\Big(1-\dfrac{i\bar \nu({\bf k,q})\cdot
\alpha_{\omega,\bar\nu}(\mathbf{q})}{\omega+i\bar \nu({\bf k,q})}\Big)+
$$
$$
+\dfrac{f_{{\bf k}+{\mathbf{q}}/{2}}-f_{{\bf k}-{\mathbf{q}}/{2}}}
{\E_{{\bf k}-{\bf q}/2}-\E_{{\bf k}+{\bf q}/2}}\cdot
\dfrac{i\bar \nu({\bf k,q})\cdot
\alpha_{\omega,\bar\nu}(\mathbf{q})}{\omega+i\bar \nu({\bf k,q})}\Bigg].
\eqno{(3.1)}
$$

Here, we will remind, that
$$
\alpha_{\omega,\bar\nu}(\mathbf{q})=
\dfrac{B_{\bar\nu}({\bf q},\omega+i\bar \nu)}{B_{\bar\nu}({\bf q},0)-
iB_{\omega,\bar\nu}({\bf q},0)+iB_{\omega,\bar\nu}({\bf q},\omega+i\bar \nu)}.
$$\medskip

Density of current ${\bf j}_e({\bf q}, \omega,\bar \nu)$ is calculated through
$\langle{\bf k}_1|\rho_1|{\bf k}_2\rangle$ \medskip
$$
{\bf j}_e({\bf q}, \omega,\bar \nu)=e
{\bf j}({\bf q}, \omega,\bar \nu)=\dfrac{e\hbar}{m}
\int \dfrac{{\bf k}d\,{\bf k}}{4\pi^3}
\left\langle{\bf k}+\dfrac{\mathbf{q}}{2}\Big|\rho_1\Big|{\bf k}-
\dfrac{\mathbf{q}}{2}\right\rangle.
\eqno{(3.2)}
$$\medskip

Thus intensity of electric field is connected with potential
of this field by relation
$
{\bf E}({\bf q}, \omega)=-i{\bf q}{\varphi}({\bf q},\omega),
$
because $\varphi({\bf q},\omega)=e^{i({\bf q\,r}-\omega t)}$.

From here the field potential is equal
$$
{\varphi}({\bf q},\omega)=i\dfrac{{\bf q}{\bf E}({\bf q}, \omega)}{q^2}.
$$
Hence, expression for current density
${\bf j}_e({\bf q}, \omega,\bar \nu)$
by means of  relation (3.2) it is possible to rewrite in the form
$$
 {\bf j}_e({\bf q}, \omega,\bar \nu)=-i
\dfrac{e^2\hbar}{mq^2}{\bf E}({\bf q}, \omega)
\int \dfrac{{\bf q k}d\,{\bf k}}{4\pi^3}
R({\bf k},\mathbf{q},\omega,\bar \nu).
$$

Here according to (3.1) \medskip
$$
R({\bf k},\mathbf{q},\omega,\bar \nu)=
$$
$$=
\dfrac{f_{{\bf k}+{\mathbf{q}}/{2}}-f_{{\bf k}-{\mathbf{q}}/{2}}}
{\E_{{\bf k}-{\bf q}/2}-\E_{{\bf k}+{\bf q}/2}+
\hbar(\omega+i\bar \nu({\bf k,q}))}\Big(1-\dfrac{i\bar \nu({\bf k,q})\cdot
\alpha_{\omega,\bar\nu}(\mathbf{q})}{\omega+i\bar \nu({\bf k,q})}\Big)+
$$\medskip
$$
+\dfrac{f_{{\bf k}+{\mathbf{q}}/{2}}-f_{{\bf k}-{\mathbf{q}}/{2}}}
{\E_{{\bf k}-{\bf q}/2}-\E_{{\bf k}+{\bf q}/2}}\cdot
\dfrac{i\bar \nu({\bf k,q})\cdot
\alpha_{\omega,\bar\nu}(\mathbf{q})}{\omega+i\bar \nu({\bf k,q})}.
$$\medskip

Considering connection of density of the current with intensity of the
field, we receive expression for electric conductivity in quantum
collisional plasma
$$
\sigma_l(\mathbf{q},\omega,\bar \nu)=-i
\dfrac{e^2\hbar}{mq^2}\int \dfrac{{\bf q k}d\,{\bf k}}{4\pi^3}
R({\bf k},\mathbf{q},\omega,\bar \nu).
\eqno{(3.2)}
$$

By means of (3.2) we will write expression for dielectric permeability
$$
\varepsilon_l(\mathbf{q},\omega,\bar \nu)=1+\dfrac{4\pi e^2\hbar}
{m\omega q^2}\int \dfrac{\mathbf{k\,q}d\,{\bf k}}{4\pi^3}
R({\bf k},\mathbf{q},\omega,\bar \nu).
\eqno{(3.3)}
$$

Scalar product $ \mathbf{k \, q} $ we will express from relation (2.2)
$$
\mathbf{k\,q}=\dfrac{m}{\hbar^2}\Big(\E_{\mathbf{k}+\mathbf{q}/2}-
\E_{\mathbf{k}-\mathbf{q}/2}\Big).
$$

By means of this expression we will copy relations (3.2) and (3.3) in
the following form
$$
\sigma_l(\mathbf{q},\omega,\bar \nu)=-i\dfrac{e^2}{\hbar q^2}\int
\dfrac{d{\bf k}}{4\pi^3}K(\mathbf{k,q},\omega,\bar \nu)
\eqno{(3.4)}
$$
and
$$
\varepsilon_l(\mathbf{q},\omega,\bar \nu)=1+\dfrac{4\pi e^2}
{\hbar\omega q^2}\int \dfrac{d{\bf k}}{4\pi^3}
K(\mathbf{k,q},\omega,\bar \nu).
\eqno{(3.5)}
$$

Here
$$
K(\mathbf{k,q},\omega,\bar \nu)=R(\mathbf{k,q},\omega,\bar \nu)
\Big(\E_{\mathbf{k}+\mathbf{q}/2}-\E_{\mathbf{k}-\mathbf{q}/2}\Big).
$$

Calculating this quantity in an explicit form, we receive
$$
K(\mathbf{k,q},\omega,\bar \nu)=-(f_{{\bf k+q}/2}-f_{{\bf k-q}/2})+
$$
$$
+\hbar \dfrac{f_{{\bf k+q}/2}-f_{{\bf k-q}/2}}{\E_{{\bf k-q}/2}-\E_{{\bf k+q}/2}
+\hbar (\omega+i\bar\nu({\bf k,q}))}[\omega+i\bar\nu({\bf k,q})(1-
\alpha_{\omega,\bar\nu}(\mathbf{q}))].
$$\medskip

Let's substitute this expression in equality (3.4). We receive, that
$$
\sigma_l(\mathbf{q},\omega,\bar \nu)=-i
\dfrac{e^2}{q^2}\int\dfrac{d{\bf k}}{4\pi^3}
\dfrac{f_{{\bf k+q}/2}-f_{{\bf k-q}/2}}{\E_{{\bf k-q}/2}-\E_{{\bf k+q}/2}
+\hbar (\omega+i\bar\nu({\bf k,q}))}\times
$$
$$
\times[\omega+i\bar\nu({\bf k,q})(1-
\alpha_{\omega,\bar\nu}(\mathbf{q}))],
$$
or, that all the same,
$$
\sigma_l(\mathbf{q},\omega,\bar \nu)=-i
\dfrac{e^2}{q^2}\Big[\omega B({\bf q},\omega+i\bar\nu)+
iB_{\bar\nu}({\bf q},\omega+i\bar \nu)(1-
\alpha_{\omega,\bar\nu}(\mathbf{q}))].
\eqno{(3.6)}
$$

Here
$$
 B({\bf q},\omega+i\bar\nu)=\int \dfrac{d{\bf k}}{4\pi^3}
 \dfrac{f_{{\bf k+q}/2}-f_{{\bf k-q}/2}}{\E_{{\bf k-q}/2}-\E_{{\bf k+q}/2}
 +\hbar(\omega+i\bar \nu({\bf k,q}))}.
$$

On the basis of (3.6) we will write out expression for dielectric function
$$
\varepsilon_l(\mathbf{q},\omega,\bar \nu)=1+\dfrac{4\pi e^2}{\omega q^2}
\times $$$$ \times
\Big[\omega B({\bf q},\omega+i\bar\nu)+
iB_{\bar\nu}({\bf q},\omega+i\bar \nu)(1-\alpha_{\omega,\bar\nu}({\bf q}))].
\eqno{(3.7)}
$$

From the formula (3.7) it is visible that at $ \omega=0$ we receive
$$
\varepsilon_l(\mathbf{q},0,\nu)=1 +\dfrac{4\pi e^2}{q^2}B({\bf q},0).
$$
Thus, at $ \omega=0$ dielectric function does not depend
from frequency of particles collisions of plasma.

At $\nu=0$ from (3.7) we receive
$$
\varepsilon_l(\mathbf{q},\omega,0) =
1+\dfrac{4\pi e^2}{q^2}B({\bf q},\omega).
$$
Thus, at $ \nu=0$ dielectric function will be transformed to
the known formula received by Klimontovich and Silin in 1952
and after that by Lindhard  in 1954.

We find explicit form of expressions (3.6) and (3.7). For
quantity $1-\alpha_{\omega,\bar\nu}({\bf q})$ we have
$$
1-\alpha_{\omega,\bar\nu}({\bf q})=$$$$=\dfrac{[B_{\bar\nu}({\bf q},0)-
B_{\bar\nu}({\bf q},\omega+i\bar \nu)]-i[B_{\omega,\bar\nu}({\bf q},0)-
B_{\omega,\bar\nu}({\bf q},\omega+i\bar \nu)]}{B_{\bar\nu}({\bf q},0)-
iB_{\omega,\bar\nu}({\bf q},0)+iB_{\omega,\bar\nu}({\bf q},\omega+i
\bar\nu)}.
$$

We will notice that
$$
\dfrac{1}{\E_{{\bf k-q}/2}-\E_{{\bf k+q}/2}}-
\dfrac{1}{\E_{{\bf k-q}/2}-\E_{{\bf k+q}/2}+\hbar(\omega+i
\bar \nu(\mathbf{k,q}))}=
$$
$$
=\dfrac{\hbar(\omega+i \bar \nu(\mathbf{k,q}))}
{(\E_{{\bf k-q}/2}-\E_{{\bf k+q}/2})
(\E_{{\bf k-q}/2}-\E_{{\bf k+q}/2}+\hbar(\omega+i \bar \nu(\mathbf{k,q})))}.
$$

Hence, the first difference from numerator of the previous expression
is equal
$$
B_{\bar\nu}({\bf q},0)-B_{\bar\nu}({\bf q},\omega+i\bar \nu)=
$$
$$
=\hbar\int \dfrac{d{\bf k}}{4\pi^3}\dfrac{ \bar \nu(\mathbf{k,q})
(\omega+i \bar \nu(\mathbf{k,q}))
(f_{{\bf k+q}/2}-f_{{\bf k-q}/2})}
{(\E_{{\bf k-q}/2}-\E_{{\bf k+q}/2})(\E_{{\bf k-q}/2}-
\E_{{\bf k+q}/2}+\hbar(\omega+i \bar \nu(\mathbf{k,q})))}.
$$

The second difference from numerator of the previous expression is equal
$$
B_{\omega,\bar\nu}({\bf q},0)-B_{\omega,\bar\nu}({\bf q},\omega+i\bar \nu)=
$$
$$
=\hbar \int  \dfrac{d{\bf k}}{4\pi^3}\dfrac{ \bar \nu^2(\mathbf{k,q})
(f_{{\bf k+q}/2}-f_{{\bf k-q}/2})}
{(\E_{{\bf k-q}/2}-\E_{{\bf k+q}/2})(\E_{{\bf k-q}/2}-
\E_{{\bf k+q}/2}+\hbar(\omega+i \bar \nu(\mathbf{k,q})))}.
$$

Hence, all numerator is equal
$$
[B_{\bar\nu}({\bf q},0)-
B_{\bar\nu}({\bf q},\omega+i\bar \nu)]-i[B_{\omega,\bar\nu}({\bf q},0)-
B_{\omega,\bar\nu}({\bf q},\omega+i\bar \nu)]=
$$ \medskip
$$
=\hbar \omega\int \dfrac{d{\bf k}}{4\pi^3}\dfrac{ \bar \nu(\mathbf{k,q})
(f_{{\bf k+q}/2}-f_{{\bf k-q}/2})}
{(\E_{{\bf k-q}/2}-\E_{{\bf k+q}/2})(\E_{{\bf k-q}/2}-
\E_{{\bf k+q}/2}+\hbar(\omega+i \bar \nu(\mathbf{k,q})))}=
$$\medskip
$$
=\omega\int \dfrac{d{\bf k}}{4\pi^3}
\dfrac{ \bar \nu(\mathbf{k,q})}{\omega+i \bar\nu(\mathbf{k,q})}\cdot
\dfrac{f_{{\bf k+q}/2}-f_{{\bf k-q}/2}}{\E_{{\bf k-q}/2}-\E_{{\bf k+q}/2}}-
$$\medskip
$$
-\omega\int \dfrac{d{\bf k}}{4\pi^3}
\dfrac{ \bar \nu(\mathbf{k,q})}{\omega+i \nu(\mathbf{k,q})}\cdot
\dfrac{f_{{\bf k+q}/2}-f_{{\bf k-q}/2}}{\E_{{\bf k-q}/2}-
\E_{{\bf k+q}/2}+\hbar(\omega+i\bar \nu(\mathbf{k,q}))}=
$$\medskip
$$
=\omega[b_{\omega,\bar \nu(\mathbf{k,q})}({\bf q},0)-
b_{\omega,\bar\nu}({\bf q},\omega+i\bar \nu)],
$$
where
$$
b_{\omega,\bar \nu}({\bf q},0)=\int \dfrac{d{\bf k}}{4\pi^3}
\dfrac{ \bar \nu(\mathbf{k,q})}{\omega+i\bar \nu(\mathbf{k,q})}\cdot
\dfrac{f_{{\bf k+q}/2}-f_{{\bf k-q}/2}}{\E_{{\bf k-q}/2}-
\E_{{\bf k+q}/2}+\hbar(\omega+i \bar\nu(\mathbf{k,q}))},
$$\medskip
$$
b_{\omega,\bar\nu}({\bf q},\omega+i \nu)=
\int \dfrac{d{\bf k}}{4\pi^3}\dfrac{ \bar \nu(\mathbf{k,q})}
{\omega+i\bar \nu(\mathbf{k,q})}\cdot
\dfrac{f_{{\bf k+q}/2}-f_{{\bf k-q}/2}}{\E_{{\bf k-q}/2}-
\E_{{\bf k+q}/2}+\hbar(\omega+i\bar \nu(\mathbf{k,q}))}.
$$

Now we find  the denominator
$$
B_{\bar\nu}({\bf q},0)-
iB_{\omega,\bar\nu}({\bf q},0)+iB_{\omega,\bar\nu}({\bf q},\omega+i\bar\nu).
$$

We notice that
$$
B_{\bar \nu}({\bf q},0)-B_{\omega,\bar \nu}({\bf q},0)=
\omega\int \dfrac{d{\bf k}}{4\pi^3}\dfrac{f_{{\bf k+q}/2}-f_{{\bf k-q}/2}}
{\E_{{\bf k-q}/2}-\E_{{\bf k+q}/2}}\cdot \dfrac{\bar \nu({\bf k,q})}
{\omega+i\bar \nu({\bf k,q})}=
$$
$$
=\omega b_{\omega,\bar \nu}({\bf q},0).
$$

Hence, the denominator is equal
$$
B_{\bar\nu}({\bf q},0)-
iB_{\omega,\bar\nu}({\bf q},0)+iB_{\omega,\bar\nu}({\bf q},\omega+i\bar\nu)=
$$
$$
=\omega b_{\omega,\bar\nu}({\bf q},0)+iB_{\omega,\bar\nu}({\bf q},
\omega+i\bar \nu).
$$

Thus, we have found, that
$$
1-\alpha_{\omega,\bar\nu}({\bf q})=
\omega\dfrac{b_{\omega,\bar\nu}({\bf q},0)-b_{\omega,\bar\nu}({\bf q},
\omega+i\bar \nu)}{\omega b_{\omega,\bar\nu}({\bf q},0)+
iB_{\omega,\bar\nu}({\bf q},\omega+i\bar \nu)}.
\eqno{(3.8)}
$$

According to (3.8) electric conductivity and dielectric permeability
in quantum collisional plasma are accordingly equal to
$$
\sigma_l({\bf q},\omega,\bar\nu)=-i\dfrac{e^2}{q^2}\omega
\Big[B({\bf q},\omega+i\bar \nu)+ \hspace{5cm}
$$
$$
\hspace{3.5cm} +iB_{\bar\nu}({\bf q},\omega+i \bar\nu)
\dfrac{b_{\omega,\bar\nu}({\bf q},0)-b_{\omega,\bar\nu}({\bf q},
\omega+i \nu)}{\omega b_{\omega,\bar\nu}({\bf q},0)+
iB_{\omega,\bar\nu}({\bf q},\omega+i\bar \nu)}\Big]
\eqno{(3.9)}
$$
and
$$
\varepsilon_l({\bf q},\omega,\bar\nu)=1+\dfrac{4\pi e^2}{q^2}
\Big[B({\bf q},\omega+i\bar \nu)+ \hspace{5cm}
$$
$$
\hspace{3cm} +iB_{\bar\nu}({\bf q},\omega+i \bar\nu)
\dfrac{b_{\omega,\bar\nu}({\bf q},0)-b_{\omega,\bar\nu}({\bf q},
\omega+i\bar \nu)}{\omega b_{\omega,\bar\nu}({\bf q},0)+
iB_{\omega,\bar\nu}({\bf q},\omega+i\bar\nu)}\Big].
\eqno{(3.10)}
$$

We put $\omega=0$ in (3.10). Then
$$
\varepsilon_l({\bf q},0,\bar\nu)=1+\dfrac{4\pi e^2}{q^2}
\Big[B({\bf q},i\bar \nu)+iB_{\bar\nu}({\bf q},i \bar\nu)
\dfrac{b_{0,\bar\nu}({\bf q},0)-b_{0,\bar\nu}({\bf q},
i\bar \nu)}{iB_{0,\bar\nu}({\bf q},i\bar\nu)}\Big].
$$
We notice that
$$
b_{0,\bar\nu}({\bf q},0)=-iB({\bf q},0),\qquad
b_{0,\bar\nu}({\bf q}=-iB({\bf q},i\bar \nu),
$$
$$
B_{0,\bar\nu}({\bf q},i\bar\nu)=-iB_{\bar \nu}({\bf q},i\bar \nu).
$$

Hence,
$$
\varepsilon_l({\bf q},0,\bar\nu)=1+\dfrac{4\pi e^2}{q^2}
\Big[B({\bf q},i\bar \nu)+B_{\bar\nu}({\bf q},i \bar\nu)
\dfrac{B({\bf q},0)-B({\bf q},i\bar \nu)}{B_{\bar\nu}({\bf q},i\bar\nu)}\Big]=
$$
$$
=1+\dfrac{4\pi e^2}{q^2}B({\bf q},0)\equiv\varepsilon_l({\bf q}).
$$

\begin{center}
\bf 4. Non-degenerate plasmas with constant frequency of collisions
and comparison with  Mermin' formula
\end{center}

Let's consider now the case of constant frequency of electrons collisions
$$
 \nu(\mathbf {k}) = \nu =\const.
$$

Then $\bar\nu({\bf k}_1,{\bf k}_2)=\bar\nu({\bf k},{\bf q})=\nu=\const$.
As it was specified above,
the quantity $ \alpha_{\omega, \bar \nu}({\bf q}) $ for constant frequency
of collisions becomes the followinf form
$$
\alpha_{\omega, \bar \nu}({\bf q})=\dfrac{(\omega+i \nu)B({\bf q},\omega+i\nu)}
{\omega B({\bf q},0)+i \nu B({\bf q},\omega+i\nu)}.
$$

Besides,
$$
B({\bf q},\bar \nu,0,\omega+i\bar \nu)=\nu B({\bf q},\omega+i \nu),
$$

$$
1-\alpha_{\omega, \bar \nu}({\bf q})=
\omega\dfrac{B({\bf q},0)-B({\bf q},\omega+i \nu)}
{\omega B({\bf q},0)+i \nu B({\bf q},\omega+i \nu)}.
$$

Hence, according to (3.6) and (3.7) (or (3.9) and (3.10)) it is
received accordingly expres\-si\-ons
for electric conductivity and dielectric function
for quantum collisio\-nal plasmas \medskip
$$
\sigma_l(\mathbf{q},\omega,\nu)=-i\dfrac{e^2}{q^2}
\dfrac{\omega(\omega+i \nu)B({\bf q},\omega+i \nu)B({\bf q},0)}
{\omega B({\bf q},0)+i \nu B({\bf q},\omega+i \nu)}.
\eqno{(4.1)}
$$\medskip
$$
\varepsilon_l(\mathbf{q},\omega,\nu)=1+\dfrac{4\pi  e^2}{q^2}
\dfrac{(\omega+i \nu)B({\bf q},\omega+i \nu)B({\bf q},0)}
{\omega B({\bf q},0)+i \nu B({\bf q},\omega+i \nu)}.
\eqno{(4.2)}
$$\medskip

Let's compare the formula (4.2) to corresponding Mermin's formula (8)
of its work \cite{Mermin}.

Let's write out Mermin's formula (8) for dielectric function
\cite{Mermin}\medskip
$$
\varepsilon_l^{\rm Mermin}({\bf q},\omega,\nu)=1+
$$

$$+
\dfrac{(1+i/\omega\tau)(\varepsilon^0({\bf q},
\omega+i/\tau)-1)}{1+(i/\omega\tau)(\varepsilon^0({\bf q},\omega+i/\tau)-1)/
(\varepsilon^0({\bf q},0)-1)}.
\eqno{(4.3)}
$$\medskip

In expression (4.3) $\varepsilon^0(q,0)$ is Lindhard's dielectric function
for collisionless plasmas,
$$
\varepsilon^0({\bf q},\omega)=1+\dfrac{4\pi e^2}{q^2}B({\bf q},\omega),
$$ \medskip
$$
B({\bf q},\omega)=\int \dfrac{d{\bf p}}{4\pi^3}\dfrac{f_{{\bf p+q}/2}-
f_{{\bf p-q}/2}}{\E_{{\bf p-q}/2}-\E_{{\bf p+q}/2}+\omega},
$$\medskip
$$
B({\bf q},0)=\int \dfrac{d{\bf p}}{4\pi^3}\dfrac{f_{{\bf p+q}/2}-f_{{\bf p-q}/2}}
{\E_{{\bf p-q}/2}-\E_{{\bf p+q}/2}}.
$$\medskip

Let's transform the formula (4.3), noticing, that $1+i/\omega\tau = (\omega+i
\nu)/\omega $, to the form \medskip
$$
\varepsilon_l^{\rm Mermin}({\bf q},\omega,\nu)=1+\dfrac{(\omega+i \nu)
[\varepsilon^0({\bf q},\omega+i \nu)-1][\varepsilon^0({\bf q},0)-1]}
{\omega[\varepsilon^0({\bf q},0)-1]+ i \nu
[\varepsilon^0({\bf q},\omega+i \nu)-1]}.
\eqno{(4.4)}
$$\medskip

Let's copy Mermin's formula  (4.4) in terms of integrals $B({\bf q},\omega)$
and $B({\bf q},0)$\medskip
$$
\varepsilon_l^{\rm Mermin}({\bf q},\omega,\nu)=1+\dfrac{4\pi e^2}{q^2}
\dfrac{(\omega+i \nu)B({\bf q},\omega+i \nu)B({\bf q},0)}
{\omega B({\bf q},0)+i \nu B({\bf q},\omega+i \nu)}.
\eqno{(4.5)}
$$\medskip

The formula (4.5) in accuracy coincides with the formula (4.2).

\begin{center}
  \bf 5. Solution of kinetic equation of Vlasov---Boltzmann
\end{center}

In following two paragraphs we will deduce expression for electric
conductivity and dielectric permeability of classical non-degenerate
col\-li\-sio\-nal plasmas with any degree non-degeneracy
(for any  temperature).

We take kinetic Vlasov---Boltzmann equation  for col\-li\-sio\-nal
plasmas with any temperature
$$
\dfrac{\partial f}{\partial t}+{\bf v}\dfrac{\partial f}{\partial {\bf r}}+
e{\bf E}({\bf r},t)\dfrac{\partial f}{\partial {\bf p}}=
\nu [f_{eq}-f({\bf r,v},t)].
\eqno{(5.1)}
$$

Here $f_{eq}({\bf r,v})$ is the local equilibrium distribution
function of Fermi---Dirac (local Fermian)
$$
f_{eq}=\dfrac{1}{1+\exp\Big(\dfrac{mv^2}{2k_BT}-
\dfrac{\mu({\bf r})}{k_BT}\Big)},
\eqno{(5.2)}
$$
$k_B$ is the Boltzmann constant, $T$ is the plasmas temperature, $\nu$
is the frequency of electron collisions with plasmas particles,
${\bf p}=m{\bf v}$ is the electron momentum, $e$ is the electron charge,
$\mu({\bf r})$ is the chemical plasmas  potential.

Let's present the chemical potential in linear approximation as
$$
\mu({\bf r})=\mu+\delta \mu({\bf r}), \qquad \mu=\const.
$$

Let's spend linearization of the equations (5.1) concerning the absolute
Fermian
$$
f_0(v,\mu)=\dfrac{1}{1+\exp\Big(\dfrac{mv^2}{2k_BT}-\dfrac{\mu}{k_BT}\Big)},
$$
or
$$
f_0(P,\alpha)=\dfrac{1}{1+e^{P^2-\alpha}},
$$

Here ${\bf P}$ is the dimensionless momentum (velocity), $\alpha$
is the dimensionless (reduced) chemical potential,
$$
 \mathbf{P}=\sqrt{\beta}\mathbf{v},\qquad \alpha=\dfrac{\mu}{k_BT}.
$$

For this purpose we will present distribution function of electrons
in the form, linear on $ \delta \mu({\bf r}) $ concerning absolute Fermian
$$
f\equiv f(\mathbf{r,v},t)=f_0(v,\mu)+
\dfrac{\partial f_0}{\partial\alpha}e^{i({\bf kr}-\omega t)}\psi(\mathbf{v}).
\eqno{(5.3)}
$$

Here
$$
\dfrac{\partial f_0}{\partial\alpha}=g(P,\alpha)=
\dfrac{e^{P^2-\alpha}}
{(1+e^{P^2-\alpha})^2}.
$$

Making linearization of (5.2) on $ \delta\alpha $, we receive, that
$$
f_{eq}({\bf r,v})=f_0(v,\mu)+
\dfrac{\partial f_0}{\partial\alpha}\delta\alpha,\qquad
\delta \alpha=\dfrac{\delta \mu}{k_BT}.
\eqno{(5.4)}
$$
Besides, in linear approach a member with the self-consistent field
it is equal
$$
{\bf E(r,}t)\dfrac{\partial f}{\partial {\bf  p}}={\bf E(r},t)
\dfrac{\partial f_0}{\partial {\bf p}}=
e^{i({\bf kr}-\omega t)}
\dfrac{\partial f_0}{\partial\alpha}\dfrac{v_x}{k_BT}.
\eqno{(5.5)}
$$

Substituting (5.3) -- (5.5) in the equation (5.1), we receive the equation
сoncerning function $ \psi $ from which we find, that
$$
\psi({\bf v})=\dfrac{\delta \alpha e^{-i({\bf kr}-\omega t)}+
\dfrac{e\tau v_x}{k_BT}}{1-i \omega\tau+ik\tau v_x}.
\eqno{(5.6)}
$$

Let's find  сhange of quantity of chemical potential of plasma particles
$\delta \mu({\bf r})$ from the law of preservation of number of particles
$$
\int f(\mathbf{r,v},t)d\Omega=\int  f_{eq}(\mathbf{r,v},t)d\Omega,
\eqno{(5.7)}
$$
where
$$
d\Omega=\dfrac{(2s+1)d^3p}{(2\pi\hbar)^3},
$$
$s$ is the spin of plasmas particles (electrons), $s={1}/{2}$.

The equation (5.7) will be transformed to the form
$$
e^{i({\bf kr-\omega t})}\int \dfrac{\partial f_0}{\partial\alpha} \psi({\bf v})
d\Omega= \delta \alpha\int \dfrac{\partial f_0}{\partial\alpha}d\Omega,
\eqno{(5.8)}
$$
or
$$
e^{i({\bf kr-\omega t})}\int \dfrac{\partial f_0}{\partial\alpha} \psi({\bf v})
d^3v= \delta \alpha\int \dfrac{\partial f_0}{\partial\alpha}d^3v,
\eqno{(5.8')}
$$

From equation (5.8) we receive
$$
\delta \alpha=\dfrac{e^{i({\bf kr-\omega t})}}{b_0}\int
\dfrac{\partial f_0}{\partial\alpha}\psi({\bf v})d^3v.
\eqno{(5.9)}
$$

Here
$$
b_0=\int \dfrac{\partial f_0}{\partial\alpha}d^3v=4\pi g_2(\alpha),\qquad
g_2(\alpha)=\int\limits_{0}^{\infty}g(P,\alpha)P^2dP.
$$

It is easy to see, that
$$
b_0=2\pi f_0(\alpha),\qquad f_0(\alpha)=\int\limits_{0}^{\infty}
\dfrac{dP}{1+e^{P^2-\alpha}}=\int\limits_{0}^{\infty}f_0(P,\alpha)dP.
$$

Now we will substitute (5.6) in (5.9). We will have
$$
e^{-i({\bf kr-\omega t})}\delta\alpha=\dfrac{1}{b_0}
\int \dfrac{\partial f_0}{\partial \alpha}\cdot
\dfrac{\delta \alpha e^{-i({\bf kr-\omega t})}+\dfrac{e\tau v_x}{k_BT}}
{1-i\omega\tau+ik\tau v_x}d^3v.
\eqno{(5.10)}
$$

From equation (5.10) we obtain
$$
e^{-i({\bf kr-\omega t})}\delta \alpha=\dfrac{e \tau}{k_BT}
\dfrac{B_1}{b_0-B_0}=\dfrac{e \tau}{k_BT}\dfrac{B_1/b_0}{1-B_0/b_0}.
\eqno{(5.11)}
$$

Here
$$
B_0=\int\dfrac{\partial f_0}{\partial\alpha}\dfrac{d^3v}
{1-i\omega\tau+ik\tau v_x},
$$
$$
B_1=\int
\dfrac{\partial f_0}{\partial\alpha}\dfrac{v_xd^3v}{1-i\omega\tau+ik\tau v_x}.
$$

Thus, function $ \psi $ is constructed

$$
\psi({\bf v})=\dfrac{e\tau}{k_BT}\dfrac{\dfrac{B_1}{b_0-B_0}+v_x}
{1-i\omega\tau+ik\tau v_x}.
\eqno{(5.12)}
$$\medskip

\begin{center}
  \bf 6. Electric conductivity and dielectric permeability
\end{center}

From definition of density of a current follows, that
$$
{\bf j}=\sigma_l e^{i({\bf kr}-\omega t)}=e\int {\bf v}fd\Omega=
e\int v_x e^{i({\bf kr}-\omega t)} \dfrac{\partial f_0}{\partial\alpha}
\psi({\bf v})d\Omega.
\eqno{(6.1)}
$$

From this for electriv conductivity we receive
$$
\sigma_l=e\int v_x \dfrac{\partial f_0}{\partial\alpha}
\psi({\bf v})d\Omega.
\eqno{(6.2)}
$$

In more details
$$
\sigma_l=\dfrac{e^2\tau}{k_BT}
\Bigg[\int \dfrac{v_x\dfrac{\partial f_0}{\partial\alpha}d\Omega}
{1-i\omega\tau+ik\tau v_x}\cdot \dfrac{B_1}{b_0-B_0}+\int
\dfrac{v_x^2\dfrac{\partial f_0}{\partial\alpha}d\Omega}
{1-i\omega\tau+ik\tau v_x}
\Bigg],
\eqno{(6.3)}
$$
or
$$
\sigma_l=\dfrac{e^2\tau 2m^3}{k_BT(2\pi\hbar)^3}
\Big[\dfrac{B_1^2}{b_0-B_0}+B_2\Big],
\eqno{(6.4)}
$$
where
$$
B_2=\int
\dfrac{v_x^2\dfrac{\partial f_0}{\partial\alpha}d^3v}{1-i\omega\tau+ik\tau v_x}.
$$

We notice that
$$
B_1=\dfrac{b_0}{ik\tau}-\dfrac{1-i\omega\tau}{ik\tau}B_0.
$$

$$
B_2=-\dfrac{1-i\omega\tau}{ik\tau}B_1.
$$

Taking into account this relation expression (2.4) will be copied in
the form
$$
\sigma_l=\dfrac{e^2\tau 2m^3}{k_BT(2\pi\hbar)^3}B_1
\Big[\dfrac{B_1}{b_0-B_0}-\dfrac{1-i\omega\tau}{ik\tau}\Big],
\eqno{(6.5)}
$$

Now it is necessary to find expression in the square bracket from (6.5)
$$
\dfrac{B_1}{b_0-B_1}-\dfrac{1-i\omega\tau}{ik\tau}=\dfrac{1}{1-\dfrac{B_0}{b_0}}
\Big[\dfrac{B_1}{b_0}-\dfrac{1-i\omega\tau}
{ik\tau}\Big(1-\dfrac{B_0}{b_0}\Big)\Big]=$$$$=\dfrac{1}{1-\dfrac{B_0}{b_0}}
\cdot\dfrac{\omega}{k}.
$$

Hence, (6.5) looks like
$$
\sigma_l=\dfrac{e^2\tau 2m^3}{k_BT(2\pi\hbar)^3}\cdot
\dfrac{B_1}{1-\dfrac{B_0}{b_0}}
\cdot\dfrac{\omega}{k}.
\eqno{(6.6)}
$$

Let's replace here $B_1$ according to previous and we will rewrite (2.6) in
equivalent kind
$$
\sigma_l=\dfrac{e^2\tau 2m^3}{k_BT(2\pi\hbar)^3}\cdot
\dfrac{\dfrac{b_0}{ik\tau}-\dfrac{1-i\omega\tau}{ik\tau}B_0}
{1-\dfrac{B_0}{b_0}}\cdot\dfrac{\omega}{k}.
\eqno{(6.7)}
$$

We notice that
$$
\dfrac{b_0}{ik\tau}-\dfrac{1-i\omega\tau}{ik\tau}B_0=
\dfrac{1}{ik\tau}\Bigg[\int g(v,\alpha)d^3v-\dfrac{1-i\omega\tau}{ik\tau}
\int \dfrac{g(v,\alpha)d^3v}{v_x-\dfrac{\omega+i \nu}{k}}\Bigg],
$$
$$
1-\dfrac{B_0}{b_0}=1+i\dfrac{1}{2\pi f_0(\alpha)k\tau v_T}\int
\dfrac{g(P)d^3P}{P_x-z'/k},
$$
where
$$
{\bf P}=\dfrac{{\bf v}}{v_T},\qquad v_T=\dfrac{1}{\sqrt{\beta}},\qquad
z'=\dfrac{\omega+i \nu}{v_T}.
$$

By means of last relation expression of the electric
conductivity (6.7) assumes the following form

$$
\sigma_l=-i\dfrac{e^2\tau 2m^3\omega 2\pi v_T^3f_0(\alpha)}
{k_BT(2\pi\hbar)^3 k^2\tau}\times \hspace{5cm}
$$

$$
\hspace{5cm} \times\dfrac{\displaystyle 1+\dfrac{z'}{2\pi f_0(\alpha)k}\int
\dfrac{g(P,\alpha)d^3P}{P_x-z'/k}}{\displaystyle 1+\dfrac{i \nu}{2\pi
f_0(\alpha)v_Tk}\int \dfrac{g(P,\alpha)d^3P}{P_x-z'/k}}.
\eqno{(6.8)}
$$\medskip

It is easy to calculate, that numerical density (concentration)
of non-degenerate plasmas it is equal
$$
N=\dfrac{f_2(\alpha)}{\pi^2}k_T^3, \qquad k_T=\dfrac{mv_T}{\hbar},
\qquad f_2(\alpha)=\int\limits_{0}^{\infty}\dfrac{P^2dP}{1+e^{P^2-\alpha}}.
$$
$k_T$ is the thermal wave number, $v_T$ is the thermal velocity of electrons.

Expression (6.8) we will transform to the following form

$$
\dfrac{\sigma_l}{\sigma_0}=
-i\dfrac{\nu \omega f_0(\alpha)}{k^2v_T^2f_2(\alpha)}
\dfrac{\displaystyle 1+\dfrac{z'}{2\pi f_0(\alpha)k}\int
\dfrac{g(P,\alpha)d^3P}{P_x-z'/k}}{\displaystyle 1+\dfrac{i \nu}{2\pi
f_0(\alpha)v_Tk}\int \dfrac{g(P,\alpha)d^3P}{P_x-z'/k}}.
\eqno{(6.9)}
$$\medskip

In considering of $\sigma_0=\dfrac{\omega_p^2}{4\pi}\tau$, where
$\omega_p^2=\dfrac{4\pi e^2N}{m}$, $\omega_p$ is the  plasma
(Langmuir) frequency, on basis (6.9) we receive the following expression
for dielectric function
$$
\varepsilon_l=1+\dfrac{\omega_p^2f_0(\alpha)}{k^2v_T^2f_2(\alpha)}
\dfrac{\displaystyle 1+\dfrac{z'}{2\pi f_0(\alpha)k}\int
\dfrac{g(P,\alpha)d^3P}{P_x-z'/k}}{\displaystyle 1+\dfrac{i \nu}{2\pi
f_0(\alpha)v_Tk}\int \dfrac{g(P,\alpha)d^3P}{P_x-z'/k}}.
\eqno{(6.10)}
$$ \medskip

We rewrite the formula (6.10) in the following form
$$
\varepsilon_l=1+\dfrac{x_p^2}{q^2}\cdot\dfrac{f_0(\alpha)}{f_2(\alpha)}
\cdot\dfrac{1+(z/q)b_0(z/q)}{1+(iy/q)b_0(z/q)},
$$
where
$$
z=x+iy=\dfrac{\omega+i \nu}{k_Tv_T}, \qquad
b_0(z/q)=\dfrac{1}{2f_0(\alpha)}\int\limits_{-\infty}^{\infty}
\dfrac{f_0(\mu,\alpha)d\mu}{\mu-z/q}, \qquad q=\dfrac{k}{k_T}.
$$

\begin{center}
  {\bf 7.  Connection of characteristics of quantum and classical plasma}
\end{center}

Let's show, that the basic characteristics of plasma, such, as
electric conductivity and dielectric permeability
of quantum collisional non-degenerate plasmas, pass in
limit, when wave number $k $ (or Planck's constant)
tends to zero, in corresponding characteristics non-degenerate
classical collisional plasmas.

The proof we will spend for electric conductivity. We take
expression (4.1) for electric conductivity
$$
\sigma_l(\mathbf{k},\omega,\nu)=-i\dfrac{e^2}{q^2}
\dfrac{\omega(\omega+i \nu)B({\bf k},\omega+i \nu)B({\bf k},0)}
{\omega B({\bf k},0)+i \nu B({\bf k},\omega+i \nu)}.
\eqno{(7.1)}
$$

Here
$$
B({\bf k},\omega+i \nu)=\int \dfrac{d{\bf k'}}{4\pi^3}\dfrac{f_{{\bf k'+k}/2}-
f_{{\bf k'-k}/2}}{\E_{{\bf k'-k}/2}-\E_{{\bf k'+k}/2}+\hbar(\omega+i \nu)},
$$
$$
B({\bf k},0)=\int \dfrac{d{\bf k'}}{4\pi^3}\dfrac{f_{{\bf k'+k}/2}-
f_{{\bf k'-k}/2}}{\E_{{\bf k'-k}/2}-\E_{{\bf k'+k}/2}},
$$
$$
f_{{\bf k'\pm k}/2}=\dfrac{1}
{1+\exp\Big(\dfrac{\E_{{\bf k'\pm k}/2}-\mu}{k_BT}\Big)},\qquad
\E_{{\bf k'\pm k}/2}=\dfrac{\hbar^2}{2m}\Big({\bf k'}\pm\dfrac{{\bf k}}{2}\Big)^2,
$$
$$
\E_{{\bf k'+k}/2}-\E_{{\bf k'-k}/2}=\dfrac{\hbar^2}{m}{\bf k'k}=
\dfrac{\hbar^2}{m}{k_x' k}.
$$

We linearize functions $f _ {{\bf k'\pm k}/2} $ on a wave vector.
We receive, that
$$
f_{{\bf k'\pm k}/2}=f_0({\bf k'},\alpha)\mp g({\bf k'},\alpha)
\dfrac{\hbar^2{\bf k'k}}{2mk_BT},
$$
where
$$
f_0({\bf k'},\alpha)=\dfrac{1}{1+\exp\Big(\dfrac{\hbar^2{\bf k'}^2}{2mk_BT}-
\alpha\Big)},$$$$
g({\bf k'},\alpha)=\dfrac{\exp\Big(\dfrac{\hbar^2{\bf k'}^2}{2mk_BT}-\alpha\Big)}
{\Big[1+\exp\Big(\dfrac{\hbar^2{\bf k'}^2}{2mk_BT}-\alpha\Big)\Big]^2}.
$$

Therefore,
$$
f_{{\bf k'+ k}/2}-f_{{\bf k' -k}/2}=-g({\bf k'},\alpha)\dfrac{\hbar^2{k_x'k}}
{mk_BT}
$$

By means of these relations we will present integrals
$B({\bf k}, \omega+i\nu) $ in the following form
$$
B({\bf k},\omega+i\nu)=\dfrac{1}{4\pi^3k_BT}\int
\dfrac{g({\bf k'},\alpha)k'_xd^3k'}{k'_x-\dfrac{(\omega+i \nu)m}{\hbar k}}.
\eqno{(7.2)}
$$

We take the dimensionless variable of integration
$$
{\bf P}=\dfrac{\hbar {\bf k'}}{\sqrt{2mk_BT}}=\dfrac{\hbar {\bf k'}}{mv_T}=
\dfrac{{\bf k'}}{k_T},\qquad k_T=\dfrac{mv_T}{\hbar}.
$$

Then
$$
B({\bf q},\omega+i \nu)=\dfrac{k_T^3}{4\pi^3k_BT}
\int\dfrac{g(P,\alpha)P_xd^3P}{P_x-z/q},
\eqno{(7.3)}
$$
where
$$
g(P,\alpha)=\dfrac{e^{P^2-\alpha}}{\big(1+e^{P^2-\alpha}\big)^2},\qquad
z=\dfrac{\omega+i \nu}{v_Tk_T}=x+iy,\qquad q=\dfrac{k}{k_T}.
$$

It is easy to see, that expression (7.3) is equal
$$
B({\bf q},\omega+i \nu)=\dfrac{k_T^3f_0(\alpha)}{2\pi^2k_BT}+
\dfrac{k_T^3z}{4\pi^3k_BTq}\int \dfrac{g(P,\alpha)d^3P}{P_x-z/q},
$$
or, that all the same,
$$
B({\bf q},\omega+i \nu)=\dfrac{Nf_0(\alpha)}{2k_BTf_2(\alpha)}\Bigg[
1+\dfrac{z}{2\pi f_0(\alpha)q}\int \dfrac{g(P,\alpha)d^3P}{P_x-z/q}\Bigg].
\eqno{(7.4)}
$$

From (7.4) it is clear, that
$$
B({\bf q},0)=\dfrac{Nf_0(\alpha)}{2k_BTf_2(\alpha)}.
\eqno{(7.5)}
$$

Substituting (7.4) and (7.5) in (7.1), we receive expression of the longitudinal
electric conductivity in quantum collisional non-degenerate  plasma
in limit, when $k \to 0$ (or $ \hbar\to 0$)
$$
\dfrac{\sigma_l}{\sigma_0}=-i\dfrac{\omega \nu f_0(\alpha)}
{q^2v_T^2k_T^2f_2(\alpha)}\cdot \dfrac{\displaystyle 1+
\dfrac{z}{2\pi f_0(\alpha)q}
\int\dfrac{g(P,\alpha)d^3P}{P_x-z/q}}{\displaystyle 1+\dfrac{iy}{2\pi
f_0(\alpha)q}\int \dfrac{g(P,\alpha)d^3P}{P_x-z/q}}.
\eqno{(7.6)}
$$

We are convinced now, that formulas (6.9) and (7.6) coincide.

\begin{center}
\bf 8. Conclusion
\end{center}

In the present work formulas for longitudinal electric conductivity
and dielectric permeability in quantum collisional
non-degenerate plasma with any degree of non-degeneracy are deduced.
The general case, when frequency of electron collisions
with plasma particles depends on their momentum is considered.
For this purpose the kinetic equation concerning a matrix of density with
integral of collisions in relaxation form in space of momentum is used.

It is shown, that when Planck's constant tends to zero,
the deduced formulas passes in corresponding formulas for classical plasma.
It is shown also, that when frequency of collisions of particles of plasma
tends to zero (plasma passes in collisionless one), the deduced formula
passes in the known Linhard's formula  received for
collisionless plasmas.

It is shown, that when frequency of collisions is a constant,
the deduced formula for dielectric permeability
passes in known Mermin's formula.


\newpage
\begin{thebibliography}{99}
\renewcommand{\baselinestretch}{0.8}

\bibitem{Mermin} {\it Mermin N. D.} { Lindhard Dielectric Functions
in the Relaxation--Time Approximation}//
Phys. Rev. B. 1970. V. 1, No. 5. P. 2362--2363.

\bibitem{Klim}{\it Klimontovich Y. and Silin V. P.} The Spectra of
Systems of Interacting Particles//
JETF (Journal Experimental Theoreticheskoi Fiziki), {\bf 23}, 151 (1952).

\bibitem{Lin} {\it Lindhard J.} On the properties of a gas of
charged particles// Kongelige Danske Videnskabernes Selskab,
Matematisk--Fysiske Meddelelser. V. 28, No. 8 (1954), 1--57.

\bibitem{Kliewer}{\it Kliewer K. L., Fuchs R.}
Lindhard Dielectric Functions with a Finite Electron Lifetime//
Phys. Rev. 1969. V. 181. No. 2. P. 552--558.


\bibitem{Das} {\it Das A. K.} The relaxation-time approximation in the RPA
dielectric formulation// J. Phys. F: Metal Phys. 1975. V. 5. November.
2035--2040.

\bibitem{Persson} {\it Persson B. N. J.} On the mathematical structure of the
Lindhard dielectric tensor// J. Phys. C: Solid St. Phys, 13 (1980), 435--439.

\bibitem{Long} {\it Latyshev A. V., Yushkanov A. A.}
Longitudinal permettivity of a quan\-tum degenerate
collisional plasma// Teor. and Mathem. Physics, {\bf 169}(3): 1739--1749
(2011).

\bibitem{Trans} {\it Latyshev A. V., Yushkanov A. A.}
Transverse Electric Conductivity in  Collisional Quantum Plasma//
Plasma Physics Reports, 2012, Vol. 38, No. 11, pp. 899--908.

\bibitem{Trans2}{\it Latyshev A. V., Yushkanov A. A.}
Transverse electric conduc\-tivity in quantum collisional
plasma in Mermin approach// arXiv:1109.6554v1 [math-ph]
29 Sep 2011.

\bibitem{Manf} {\it Manfredi G.} { How to model quantum plasmas}//
arXiv: quant - ph/0505004.

\bibitem{Anderson} {\it Anderson D., Hall B., Lisak M., and
Marklund M.}
{Statistical effects in the multistream model for quantum plasmas}//
Phys. Rev. E {\bf 65} (2002), 046417.

\bibitem{Andres}{\it Andr\'{e}s P.,de, Monreal R., and Flores F.}
{Relaxation--time effects in the transverse dielectric function and the
electromagnetic properties of metallic surfaces and small
particles}// Phys. Rev. {\bf B}. 1986. Vol. 34, No. 10, 7365--7366.

\bibitem{Shukla1} {\it Shukla P. K. and Eliasson B.}
Nonlinear aspects of quantum plasma physics//
Uspekhy Fiz. Nauk, {\bf 53}(1) 2010;[V. 180. No. 1, 55-82 (2010) (in Russian)].

\bibitem{Shukla2} {\it Eliasson B. and Shukla P.K.}
Dispersion properties of electrostatic oscillations in quantum
plasmas//
arXiv:0911.4594v1 [physics.plasm-ph] 24 Nov 2009, 9 pp.

\bibitem{BGK}{\it Bhatnagar P. L., Gross E. P., and Krook M.}
{A model for collision processes in gases. I. Small
amplitude processes in charged and neutral one-component systems}//
Phys. Rev. {\bf 94} (1954), 511--525.

\bibitem{Opher}{\it Opher M., Morales G. J., Leboeuf J. N.}
Krook collisional models of the kinetic susceptibility of plasmas//
Phys. Rev. E. V.66, 016407, 2002.

\bibitem{Gelder} {\it Gelder van, A. P.} Quantum Corrections in the
Theory of the Anomalous Skin Effect//
Phys. Rev. 1969. Vol. 187. No. 3. P. 833--842.

\bibitem{Fuchs}{\it Fuchs R., Kliewer K. L.}
Surface plasmon in a semi--infinite free--electron gas//
Phys. Rev. B. 1971. V. 3. No. 7. P. 2270--2278.

\bibitem{Fuchs2}{\it Fuchs R., Kliewer K. L.}
Optical properties of an
electron gas: further studies of a nonlocal description//
Phys. Rev. 1969. V. 185. No. 3. P. 905--913.

\bibitem{Dressel}{\it Dressel M., Gr\"{u}ner G.}
{Electrodynamics of Solids. Optical Properties of
Electrons in Matter}. - Cambridge. Univ. Press. 2003. 487 p.

\bibitem{Wier} {\it Wierling A.} {Interpolation between local
field corrections and the Drude model by a generalized Mermin
approach}//
arXiv:0812.3835v1 [physics.plasm-ph] 19 Dec 2008.

\bibitem{Brod} {\it Brodin G., Marklund M., Manfredi G.}
{Quantum Plasma Effects in the Classical Regime}//
Phys. Rev. Letters. {\bf 100}, (2008). P. 175001-1--175001-4.

\bibitem{Manf2} {\it Manfredi G. and Haas F.}
{Self-consistent fluid model for a quantum electron gas}//
Phys. Rev. B {\bf 64} (2001), 075316.

\bibitem{Wigner} {\it Wigner E. P.}
{On the quantum correction for thermodynamic equilibrium}//
Phys. Rev. {\bf 40} (1932), 749--759.

\bibitem{Tatarskii} {\it Tatarskii V. I.}
{The Wigner representation of quantum mechanics}//
Uspekhy Fiz. Nauk. {\bf 26} (1983), 311--327;
[Usp. Fis. Nauk. {\bf 139} (1983), 587 (in Russian)].

\bibitem{Hillery}{\it Hillery M., O'Connell R. F., Scully M.
O., and Wigner E. P.}
{Distribution functions in physics: Fundamentals}//
Phys. Rev. {\bf 106} (1984), 121--167.

\bibitem{Ropke} {\it Reinholz H., R\"{o}pke G.} Dielectric function beyond
the random-phase approximation: Kinetic theory versus linear response
theory// Phys. Rev., {\bf E 85}, 036401 (2012).

\end{thebibliography}
\end{document}